\pdfoutput=1
\documentclass[a4paper,11pt]{article}
\usepackage{paper}
\usepackage[T1]{fontenc}
\usepackage{newtxtext,newtxmath}
\usepackage{bm,booktabs,graphicx}
\usepackage[section]{placeins}
\usepackage[hidelinks]{hyperref}
\setkeys{Gin}{width=\textwidth,height=0.45\textheight,keepaspectratio}
\newcommand{\T}{\top}
\newcommand{\dd}{\mathrm{d}}
\newcommand{\ms}{\,\mathrm{m\,s^{-1}}}
\title{Dynamic conditional density network of wind vectors}
\author{Keigo Sato\thanks{E-mail: \href{mailto:sato.keigo.s7@dc.tohoku.ac.jp}{sato.keigo.s7@dc.tohoku.ac.jp}}%
  \hphantom{\textsuperscript{$\ast$}}\ and Yasumasa Matsuda\\[0.5em]
  \small Graduate School of Economics and Management, Tohoku University, Japan}
\date{}
\begin{document}
\maketitle

\begin{abstract}
We propose the dynamic conditional density network (DCDN) to correct the bivariate normal predictive distribution commonly used to postprocess ensemble forecasts of wind vectors.
The conditional distribution of the observed wind vector given the ensemble forecast may depart from the bivariate normal distribution.
DCDN multiplies a bivariate normal reference density by a correction density.
The correction density is a B-spline mixture on the probability integral transform scale of the reference density, and a neural network predicts its weights.
DCDN equals its reference density when the correction density is uniform.
We apply DCDN to ensemble forecasts of the European Centre for Medium-Range Weather Forecasts and observations at 154 sites in the Tohoku region of Japan.
DCDN significantly improved the logarithmic score, energy score, and variogram score over the bivariate normal distribution predicted by a neural network.
The improvement in the logarithmic score was large near zero wind speed and at higher wind speeds.
This is consistent with non-normal features of the conditional distribution that the bivariate normal distribution misses.
\end{abstract}
\keywords{wind vector; ensemble postprocessing; distributional regression network; B-spline mixtures; probabilistic forecasting}

\section{Introduction}\label{sec:intro}

Weather forecasts are based on numerical weather prediction (NWP) models.
Ensemble prediction systems address the uncertainty of NWP forecasts by generating a set of forecasts, called an ensemble, from perturbed initial conditions or model physics.
Ensemble forecasts, however, tend to be biased and underdispersed, in that the ensemble spread is too small to be realistic \citep{schuhen2012ensemble}.
Statistical postprocessing corrects these errors by transforming the ensemble into a predictive distribution \citep{gneiting2005calibrated}.

For wind vectors, postprocessing has used the bivariate normal distribution as the predictive distribution.
Probabilistic wind forecasts are important for applications such as air traffic control, ship routing, and wind energy \citep{schuhen2012ensemble}.
\citet{schuhen2012ensemble} proposed bivariate ensemble model output statistics (BIV-EMOS), and \citet{lang2019bivariate} proposed a distributional regression model, both based on the bivariate normal distribution.
\citet{buchner2026bivariate} proposed the bivariate distributional regression neural network (BIV-DRN) with the same distribution.

The bivariate normal assumption does not always match the conditional distribution of the observed wind vector given the forecast.
A bivariate normal distribution is fully determined by its mean and covariance, and its density is necessarily unimodal and elliptically symmetric.
However, \citet{schuhen2012ensemble} suggested that the conditional distribution of the observed wind vector, given the ensemble forecast, tends to be skewed.
The normal margins and the Gaussian dependence implied by the bivariate normal distribution might also not be justified by the data at all sites \citep{buchner2026bivariate}.

In a recent comparison for wind vectors, BIV-DRN had better scores than more flexible methods \citep{buchner2026bivariate}.
The flexible methods combine kernel density estimates of the margins with pair copulas, instead of the bivariate normal distribution.
Over 60 sites, BIV-DRN had the best logarithmic score (LogS), energy score (ES), and variogram score (VS) among the compared methods.
The flexible methods were nevertheless better calibrated than the methods based on the bivariate normal distribution.

In contrast to these methods, we propose the dynamic conditional density network (DCDN), which corrects a bivariate normal distribution instead of replacing it.
DCDN multiplies a bivariate normal reference density, here that of BIV-DRN, by a correction density.
The correction density is defined on the probability integral transform (PIT) scale of the reference density.
Following \citet{matsuda2025density}, it is a mixture of tensor products of normalized B-splines.
A neural network predicts its mixture weights for each forecast.
When the correction density is uniform, the predictive distribution of DCDN equals the bivariate normal reference distribution.

Over the test period of 2025, DCDN significantly improved the LogS, the ES, and the VS over BIV-DRN.
We compare DCDN with the raw ensemble forecasts (Raw ensemble), BIV-EMOS, and BIV-DRN.
The data are ensemble forecasts of the European Centre for Medium-Range Weather Forecasts and observations at 154 sites in the Tohoku region of Japan.
Relative to BIV-EMOS, DCDN improved the LogS by 18.9\%, whereas BIV-DRN improved it by 6.0\%.
In the ES and the VS, the improvements were 3.3\% and 4.0\% for DCDN and 2.5\% and 3.3\% for BIV-DRN.
The improvement in the LogS was large near zero wind speed and at higher wind speeds.

\Sec{sec:data} describes the data, and \Sec{sec:method} the methods.
\Sec{sec:results} presents the results, and \Sec{sec:discussion} discusses them and concludes.

\section{Data}\label{sec:data}

\subsection{Observations}\label{sec:observations}

\begin{figure}[t]
  \centering
  \includegraphics{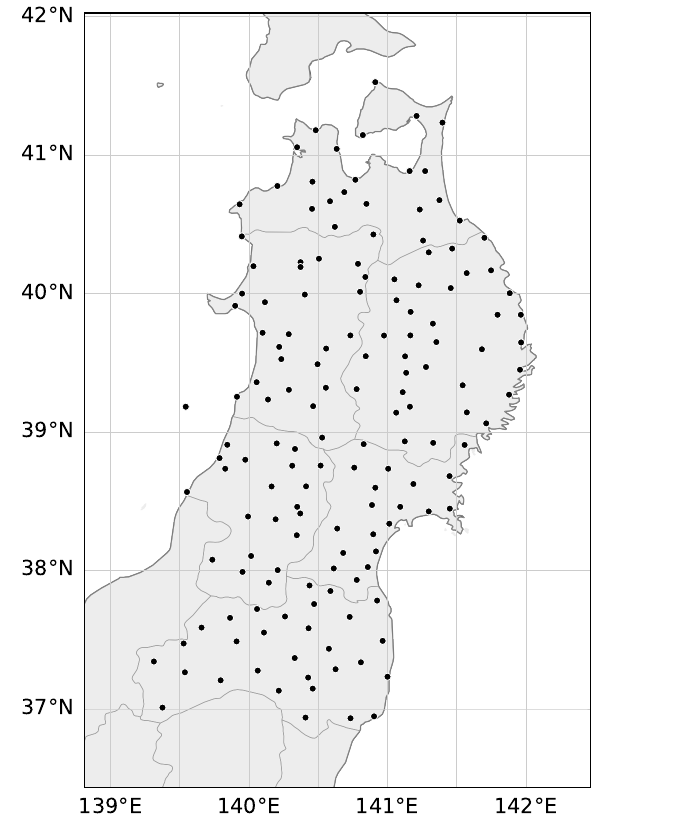}
  \caption{Locations of the 154 observation sites in the Tohoku region of Japan}
  \label{fig:sites}
\end{figure}

The observations are wind vectors recorded by the Japan Meteorological Agency (JMA) at 154 sites in the Tohoku region of Japan.
They are retrieved from the JMA website for downloading past weather data \citep{jmaobsdl}.
\Fig{fig:sites} shows the locations of the sites.
Each wind vector $\mathbf{y}$ consists of the zonal component $y_u$ and the meridional component $y_v$.

The observations are recorded as discrete values.
Wind speeds are recorded in units of $0.1\ms$, and the nonzero wind vectors point in 16 directions at intervals of $22.5^\circ$.
JMA records calm, a wind speed of $0.2\ms$ or less, without a direction \citep{jma1998guidelines}, and we treat it as the zero vector.
Because the recorded observations are discrete, training and evaluation use dequantized observations.

\label{sec:dequantization}A dequantized observation $\tilde{\mathbf{y}}_{\mathrm{obs}}$ is drawn uniformly from the set of wind vectors that would be recorded as the observation $\mathbf{y}_{\mathrm{obs}}$.
For a non-calm observation, this set is the sector within $0.05\ms$ of its speed and $11.25^\circ$ of its direction.
For a calm observation, it is the disk of radius $0.25\ms$ centered at the origin.
This radius adds half the speed interval to the upper limit of calm.
The perturbations are drawn anew for each training batch but are fixed for each validation and test case and shared by all methods.

Dequantization prevents a continuous predictive density from raising its likelihood by concentrating on the recorded values.
Without it, fitting a continuous density to discrete data produces a degenerate solution that places all probability mass on the discrete values \citep{ho2019flow}.
In density modeling of images, the discrete pixel values are therefore commonly dequantized by adding uniform noise over the width of each discrete bin \citep{ho2019flow}.
\citet{schuhen2012ensemble} replaced each wind speed observation of zero by a uniform random number between 0 and 2~kt when computing the continuous ranked probability score.
Our disk for calm extends this replacement to wind vectors.

\subsection{Ensemble forecasts}\label{sec:forecasts}

The forecasts are ensemble forecasts of the European Centre for Medium-Range Weather Forecasts (ECMWF) from the archive of TIGGE (The International Grand Global Ensemble).
TIGGE collects the ensemble predictions of global numerical weather prediction centers \citep{bougeault2010thorpex,swinbank2016tigge}.
The forecasts are retrieved on the octahedral reduced Gaussian grid O640, whose spacing is about 18~km.
They are initialized at 0000 UTC and have 60 lead times from 6 to 360~h at 6-h intervals.
Each forecast consists of the control forecast and 50 perturbed members of the zonal and meridional wind components at 10~m.

A case is a combination of a site, an initialization time, and a lead time.
For each case, the forecast is taken at the grid point nearest to the site, and the observation at the valid time.
Only the cases with an observation are used for training and evaluation.

\subsection{Training, validation, and test periods}\label{sec:periods}

The data are split into training, validation, and test periods by initialization time.
The training period is 2021--2023, the validation period is 2024, and the test period is 2025.
The validation period is used for the preliminary search of the hyperparameters and for early stopping, and the test period for the evaluation in \Sec{sec:results}.
The initialization times in the last 15 days of each period are excluded.
The valid times of the forecasts therefore fall within the period, and adjacent periods are separated by a gap of 15 days.
The validation and test periods each have about 3.2 million cases with an observation.

\section{Methods}\label{sec:method}

\subsection{Predictive distribution of DCDN}\label{sec:dcdn}

\begin{figure}[t]
  \centering
  \includegraphics{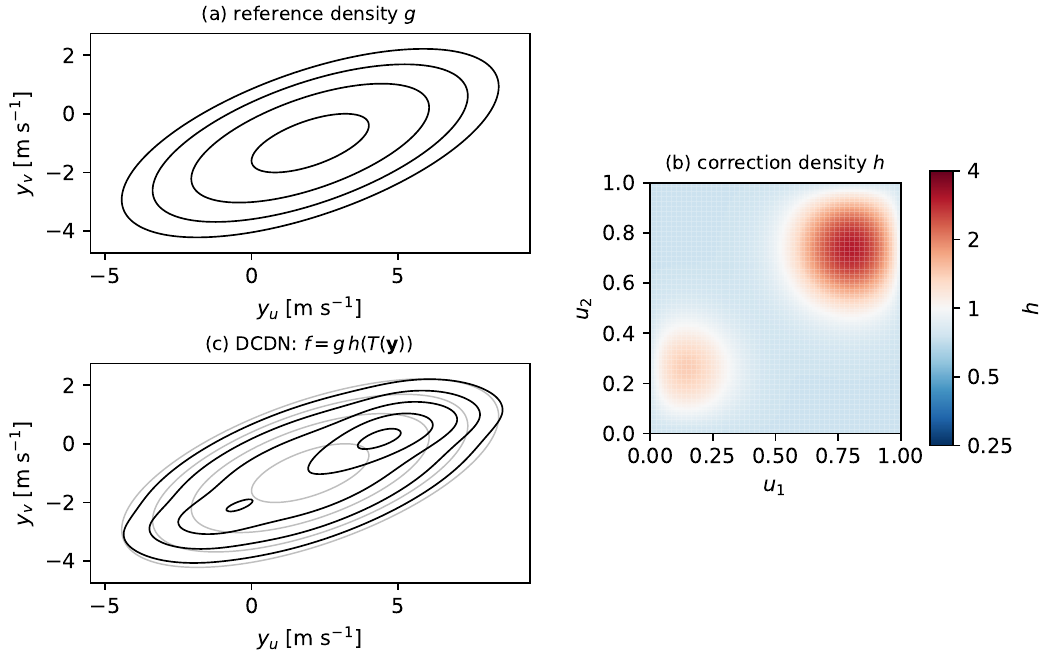}
  \caption{Schematic of DCDN. (a) Reference density $g$. (b) Correction density $h$ on the PIT scale. (c) Predictive density $f$ of DCDN, with the contours of $g$ in gray}
  \label{fig:schematic}
\end{figure}

The predictive density of DCDN is the reference density $g$ multiplied by a correction density $h$ evaluated at a transformed wind vector (\Fig{fig:schematic}):
\begin{equation}
  f(\mathbf{y} \mid \mathbf{x}) = g(\mathbf{y} \mid \mathbf{x})\, h\bigl(T(\mathbf{y}) \mid \mathbf{x}\bigr).
  \label{eq:dcdn}
\end{equation}
Here, $\mathbf{x}$ denotes the inputs of a case (\Sec{sec:networks}).
The reference density $g$ is a bivariate normal density with mean $\boldsymbol{\mu}$ and covariance
\begin{equation}
  \boldsymbol{\Sigma} = \begin{pmatrix} \sigma_u^2 & \rho\,\sigma_u \sigma_v \\ \rho\,\sigma_u \sigma_v & \sigma_v^2 \end{pmatrix},
  \label{eq:cov}
\end{equation}
where $\sigma_u$ and $\sigma_v$ are the standard deviations and $\rho$ is the correlation coefficient.
In this study, BIV-DRN gives the mean and the covariance from the inputs (\Sec{sec:networks}).

The transformation $T$ whitens the wind vector with the mean and the symmetric square root of the covariance of $g$.
It then applies the standard normal cumulative distribution function $\Phi$ to each component:
\begin{equation}
  \mathbf{z} = \boldsymbol{\Sigma}^{-1/2}(\mathbf{y} - \boldsymbol{\mu}), \qquad
  \mathbf{u} = T(\mathbf{y}) = \Phi(\mathbf{z}).
  \label{eq:transform}
\end{equation}
If $\mathbf{y}$ follows $g$, $\mathbf{u}$ is uniformly distributed on the unit square, which we call the PIT scale of the reference density.
Because the Jacobian determinant of $T$ equals $g$, $f$ is a density whenever $h$ is a density on the unit square.
In general, $h$ is the density of $T(\mathbf{y})$ when $\mathbf{y}$ follows $f$.

\subsection{Correction density}\label{sec:correction}

The correction density $h$ on the PIT scale is built from normalized cubic B-splines, following the representation of densities on the unit hypercube by \citet{matsuda2025density}.
In each coordinate, we use 13 cubic B-splines $B_i$.
Their knots have multiplicity four at 0 and 1 and are spaced 0.1 apart in between.
These B-splines are nonnegative and sum to one on the unit interval.
The normalized basis functions are
\begin{equation}
  \phi_i(u) = \frac{B_i(u)}{m_i}, \qquad m_i = \int_0^1 B_i(u)\, \dd u .
  \label{eq:basis}
\end{equation}
Each $\phi_i$ is a density on the unit interval.

The correction density is the mixture
\begin{equation}
  h(\mathbf{u} \mid \mathbf{x}) = \sum_{i=1}^{13} \sum_{j=1}^{13} p_{ij}(\mathbf{x})\, \phi_i(u_1)\, \phi_j(u_2), \qquad
  p_{ij} > 0, \qquad \sum_{i,j} p_{ij} = 1 .
  \label{eq:mixture}
\end{equation}
Because each product of two basis functions integrates to one over the unit square, $h$ is a density on the unit square.

Following \citet{matsuda2025density}, we transform the correction densities into an unconstrained Euclidean space through generalized logit and softmax mappings of the weights.
In this study, we use the weighted softmax
\begin{equation}
  p_{ij} = \frac{m_i m_j \exp\bigl((\mathbf{Q}^\T \mathbf{q})_{ij}\bigr)}{\sum_{i',j'} m_{i'} m_{j'} \exp\bigl((\mathbf{Q}^\T \mathbf{q})_{i'j'}\bigr)} ,
  \label{eq:weights}
\end{equation}
where $\mathbf{q}$ is a vector of 168 real values.
The matrix $\mathbf{Q}$ is a Helmert-type contrast matrix with 168 orthonormal rows of length 169, all orthogonal to the vector of ones.
The weights are positive and sum to one, and the mapping is inverted by the logit embedding
\begin{equation}
  \mathbf{q} = \mathbf{Q} \log\bigl(\mathbf{p} / \operatorname{vec}(\mathbf{m}\mathbf{m}^\T)\bigr),
  \label{eq:embedding}
\end{equation}
where $\mathbf{p}$ and $\mathbf{m}$ are the vectors of the weights and the integrals, $\operatorname{vec}$ follows the order of $\mathbf{p}$, and the operations are componentwise.
Weights that are positive and sum to one therefore correspond one to one to $\mathbf{q}$.
At $\mathbf{q} = \mathbf{0}$, and only there, each weight equals the product of the integrals of its two basis functions, so $h$ is one everywhere and DCDN equals the reference density.

\subsection{Networks and inputs}\label{sec:networks}

BIV-DRN and DCDN use the same inputs $\mathbf{x}$: the 16 variables listed in \Tab{tab:inputs} (\App{app:inputs}) and the site.
Following \citet{rasp2018neural}, each network represents the site by a site embedding, an entity embedding \citep{guo2016entity} that is learned with the weights and concatenated with the other inputs.

\begin{table}[t]
  \centering
  \caption{Input variables of BIV-EMOS, BIV-DRN, and DCDN. BIV-DRN and DCDN also use a site embedding}
  \label{tab:inputs}
  \begin{tabular}{lc}
    \toprule
    Input & Number \\
    \midrule
    Control forecast of each component & 2 \\
    Ensemble mean of each component & 2 \\
    Logarithm of the ensemble standard deviation of each component & 2 \\
    Ensemble skewness of each component & 2 \\
    Ensemble excess kurtosis of each component & 2 \\
    Ensemble correlation between the two components & 1 \\
    Lead time & 1 \\
    Sine and cosine of the time of day & 2 \\
    Sine and cosine of the day of the year & 2 \\
    \midrule
    Total & 16 \\
    \bottomrule
  \end{tabular}
\end{table}

BIV-DRN predicts the five parameters of the bivariate normal distribution from the inputs with a fully connected neural network.
\citet{buchner2026bivariate} extended the univariate distributional regression network to this bivariate setting for wind vectors.
Our BIV-DRN differs from theirs in the inputs and the number of hidden layers, and it is a single network, not an average of ten.
Its hidden layers use the rectified linear unit (ReLU) as the activation function, and \Sec{sec:training} gives their number and width.
The five outputs are the two means, the logarithms of the two standard deviations, and the transformed correlation coefficient
\begin{equation*}
  \tilde\rho = \frac{\rho}{\sqrt{1 - \rho^2}}, \qquad
  \rho = \frac{\tilde\rho}{\sqrt{1 + \tilde\rho^2}} ,
\end{equation*}
as in \citet{buchner2026bivariate}.

The correction network of DCDN predicts the 168 values $\mathbf{q}$ from the inputs.
Like BIV-DRN, it is fully connected with ReLU in the hidden layers (\Sec{sec:training}).
Its site embedding is separate from that of BIV-DRN.
The output layer is initialized at zero, so training of the correction network starts from $\mathbf{q} = \mathbf{0}$.

\subsection{Training}\label{sec:training}

Training proceeds in two stages: BIV-DRN is trained first, and the correction network is then trained with BIV-DRN fixed as the reference density.
In both stages, the loss is the negative log-likelihood of the dequantized observations.
The 16 inputs are standardized with their means and standard deviations over the training period.
The networks are implemented in PyTorch, and their weights use its default initialization unless stated otherwise.
They are trained with Adam \citep{kingma2015adam}.
Because the reference density is fixed, any difference between DCDN and BIV-DRN arises from the correction density $h$ alone.

The hyperparameters were fixed on the basis of a preliminary search over the validation period.
Both BIV-DRN and the correction network of DCDN have three hidden layers of width 256 and a learning rate of $10^{-3}$.
The embedding length is 10, and a batch consists of all cases of 32 initialization times.
The maximum number of epochs is 200, and the early-stopping patience is 10 epochs.
Early stopping monitors the LogS (\Sec{sec:scores}) over the validation period, and we use the weights from the epoch with the lowest value.

\subsection{Compared methods}\label{sec:benchmarks}

We compare four methods: Raw ensemble, BIV-EMOS, BIV-DRN, and DCDN.
Raw ensemble consists of the 50 perturbed members.
BIV-EMOS gives the five outputs of BIV-DRN (\Sec{sec:networks}) as linear functions of the 16 inputs in \Tab{tab:inputs}, with site-specific coefficients.
These outputs are the two means, the logarithms of the two standard deviations, and the transformed correlation coefficient, as in BIV-DRN.
This follows ensemble model output statistics (EMOS), a postprocessing technique based on multiple linear regression \citep{gneiting2005calibrated}.
The coefficients start at zero and are trained with the same loss, learning rate, and other training settings as BIV-DRN (\Sec{sec:training}).

Our BIV-EMOS differs from that of \citet{schuhen2012ensemble} in the predictors and the estimation method.
Its predictors are the 16 inputs rather than the ensemble means, variances, and mean wind direction.
Its coefficients are estimated jointly rather than in three phases.

\subsection{Scores}\label{sec:scores}

We evaluate the forecasts with the LogS, the ES, and the VS, which are proper scoring rules relative to suitable classes of distributions \citep{gneiting2008assessing,scheuerer2015variogram}.
The three scores are evaluated at the dequantized observation $\tilde{\mathbf{y}}_{\mathrm{obs}}$ (\Sec{sec:dequantization}).
All three are oriented as penalties, so a lower score indicates a better forecast.

The LogS uses the predictive density, namely $f$ for DCDN and the bivariate normal densities for BIV-EMOS and BIV-DRN:
\begin{equation}
  \mathrm{LogS} = -\ln f(\tilde{\mathbf{y}}_{\mathrm{obs}} \mid \mathbf{x}).
  \label{eq:logs}
\end{equation}
This is the logarithmic score \citep{gneiting2008assessing} with the natural logarithm.

The ES and the VS use samples $\mathbf{y}_1, \dots, \mathbf{y}_m$ to approximate expectations under the predictive distribution.
For the three postprocessing methods, the samples are $m = 1000$ draws from the predictive distribution (\App{app:sampling}).
For Raw ensemble, the samples are its $m = 50$ members, and only the ES and the VS are reported.
The ES is
\begin{equation}
  \mathrm{ES} = \frac{1}{m} \sum_{k=1}^{m} \bigl\lVert \mathbf{y}_k - \tilde{\mathbf{y}}_{\mathrm{obs}} \bigr\rVert - \frac{1}{2m^2} \sum_{k=1}^{m} \sum_{l=1}^{m} \bigl\lVert \mathbf{y}_k - \mathbf{y}_l \bigr\rVert,
  \label{eq:es}
\end{equation}
which is the energy score \citep{gneiting2008assessing}, a multivariate generalization of the continuous ranked probability score, with the Euclidean norm and equal masses on the samples.
The VS is
\begin{equation}
  \mathrm{VS} = 2 \Bigl( \lvert \tilde y_{\mathrm{obs},u} - \tilde y_{\mathrm{obs},v} \rvert^{0.5} - \frac{1}{m} \sum_{k=1}^{m} \bigl\lvert y_{k,u} - y_{k,v} \bigr\rvert^{0.5} \Bigr)^2 ,
  \label{eq:vs}
\end{equation}
which is the variogram score of \citet{scheuerer2015variogram} of order 0.5 with all weights equal to one.

\subsection{Comparison and tests}\label{sec:dm}

We show the improvements as percentage improvements over BIV-EMOS.
Let $S_m$ and $S_{\mathrm{EMOS}}$ be a score averaged over all cases for method $m$ and for BIV-EMOS.
The improvement of method $m$ is
\begin{equation*}
  100 \bigl\{ \exp(S_{\mathrm{EMOS}} - S_m) - 1 \bigr\} \ \text{for the LogS}
  \quad \text{and} \quad
  100 \, (1 - S_m / S_{\mathrm{EMOS}}) \ \text{for the ES and the VS},
\end{equation*}
and a positive value means that method $m$ is better than BIV-EMOS.
For the ES and the VS, the improvement is the skill score \citep{gneiting2007strictly} in percent, computed against BIV-EMOS with an optimal score of zero.
In the LogS, the improvement is the percentage by which the geometric mean of the predictive density at the observations exceeds that of BIV-EMOS.

We repeat the training of BIV-EMOS, BIV-DRN, and DCDN ten times with different random numbers (\App{app:training}).
The overall scores and improvements are averaged over the ten training runs.
Tables~\ref{tab:overall} and~\ref{tab:calm} also give their standard deviations.
Raw ensemble is not trained, so its ES and VS are single values.
The DM tests and the results shown in figures use the model from one of the ten training runs of each method.

For each of the three scores, a two-sided Diebold--Mariano (DM) test compares the three pairs among BIV-EMOS, BIV-DRN, and DCDN.
The test uses the differences in the score between two methods, averaged over all sites and lead times for each initialization time.
Let $d_t$ be this difference for the $t$th of the $N$ initialization times, and let $\bar d$ be their mean.
The test statistic is
\begin{equation}
  \mathrm{DM} = \frac{\bar d}{\sqrt{\hat V / N}}, \qquad
  \hat V = \hat\gamma_0 + 2 \sum_{\tau=1}^{L} \hat\gamma_\tau, \qquad
  \hat\gamma_\tau = \frac{1}{N} \sum_{t=1}^{N-\tau} (d_t - \bar d)(d_{t+\tau} - \bar d),
  \label{eq:dm}
\end{equation}
where $\hat\gamma_\tau$ is the sample autocovariance at lag $\tau$, and $\hat V$ uses the uniform, or rectangular, lag window of \citet{diebold1995comparing}.
Because forecasts initialized within 14 days of each other have overlapping valid times, the maximum lag $L$ is 14.
The statistic is compared with the standard normal distribution at the significance level of 0.05.

\section{Results}\label{sec:results}

\subsection{Overall scores}\label{sec:overall}

Over the test period of 2025, DCDN had lower LogS, ES, and VS than BIV-DRN (\Tab{tab:overall}).
Relative to BIV-EMOS, BIV-DRN improved the LogS, the ES, and the VS by 6.00\%, 2.48\%, and 3.31\%, respectively.
DCDN improved them by 18.91\%, 3.28\%, and 4.03\%, respectively.
All nine DM tests were significant at the 0.05 level, the weakest being the difference between BIV-DRN and DCDN in the VS ($p = 0.043$).
Raw ensemble had higher ES and VS than the three postprocessing methods.
In each of the ten training runs, the improvement of DCDN exceeded that of BIV-DRN in all three scores.

The calm cases accounted for about a fifth of the differences in the mean LogS and ES between BIV-DRN and DCDN (\Tab{tab:calm}).
In the VS, they accounted for more than the whole difference.
They were only 3.21\% of the test cases.
In the non-calm cases, BIV-DRN improved the LogS, the ES, and the VS over BIV-EMOS by 5.54\%, 2.45\%, and 3.19\%, respectively.
DCDN improved them by 16.13\%, 3.11\%, and 3.03\%, respectively.
Because the subsets are defined by the observations, ranking the methods within a subset ought to be avoided \citep{lerch2017forecasters}.

\begin{table}[t]
  \centering
  \caption{Mean scores and percentage improvements over BIV-EMOS in the test period, given as the mean $\pm$ standard deviation over ten training runs. Lower part: DM tests for one training run of each method, where a positive statistic favors the second method of the pair}
  \label{tab:overall}
  \small
  \begin{tabular}{lccc}
    \toprule
    Method & LogS & ES & VS \\
    \midrule
    Raw ensemble & -- & 1.7997 & 1.0928 \\
    BIV-EMOS & $3.2953 \pm 0.0004$ & $1.2813 \pm 0.0004$ & $0.5945 \pm 0.0002$ \\
    BIV-DRN & $3.2371 \pm 0.0069$ & $1.2495 \pm 0.0020$ & $0.5748 \pm 0.0017$ \\
    DCDN & $3.1222 \pm 0.0038$ & $1.2392 \pm 0.0010$ & $0.5706 \pm 0.0008$ \\
    \midrule
    Improvement of BIV-DRN (\%) & $6.00 \pm 0.73$ & $2.48 \pm 0.17$ & $3.31 \pm 0.29$ \\
    Improvement of DCDN (\%) & $18.91 \pm 0.45$ & $3.28 \pm 0.10$ & $4.03 \pm 0.13$ \\
    \bottomrule
  \end{tabular}

  \medskip
  \begin{tabular}{llcc}
    \toprule
    Score & Pair & DM statistic & $p$-value \\
    \midrule
    LogS & BIV-EMOS -- BIV-DRN & 10.33 & $< 0.001$ \\
     & BIV-EMOS -- DCDN & 21.60 & $< 0.001$ \\
     & BIV-DRN -- DCDN & 24.71 & $< 0.001$ \\
    \midrule
    ES & BIV-EMOS -- BIV-DRN & 12.08 & $< 0.001$ \\
     & BIV-EMOS -- DCDN & 15.97 & $< 0.001$ \\
     & BIV-DRN -- DCDN & 7.17 & $< 0.001$ \\
    \midrule
    VS & BIV-EMOS -- BIV-DRN & 12.67 & $< 0.001$ \\
     & BIV-EMOS -- DCDN & 9.20 & $< 0.001$ \\
     & BIV-DRN -- DCDN & 2.03 & 0.043 \\
    \bottomrule
  \end{tabular}
\end{table}

\begin{table}[t]
  \centering
  \caption{As the upper part of \Tab{tab:overall}, separately for non-calm and calm cases}
  \label{tab:calm}
  \small
  \begin{tabular}{lccc}
    \toprule
    Method & LogS & ES & VS \\
    \midrule
    \multicolumn{4}{l}{Non-calm cases (3,121,733 cases, 96.79\%)} \\
    \midrule
    Raw ensemble & -- & 1.8241 & 1.0692 \\
    BIV-EMOS & $3.3595 \pm 0.0004$ & $1.3088 \pm 0.0005$ & $0.5891 \pm 0.0002$ \\
    BIV-DRN & $3.3055 \pm 0.0076$ & $1.2767 \pm 0.0021$ & $0.5703 \pm 0.0018$ \\
    DCDN & $3.2100 \pm 0.0045$ & $1.2682 \pm 0.0010$ & $0.5713 \pm 0.0011$ \\
    Improvement of BIV-DRN (\%) & $5.54 \pm 0.80$ & $2.45 \pm 0.17$ & $3.19 \pm 0.32$ \\
    Improvement of DCDN (\%) & $16.13 \pm 0.52$ & $3.11 \pm 0.10$ & $3.03 \pm 0.18$ \\
    \midrule
    \multicolumn{4}{l}{Calm cases (103,484 cases, 3.21\%)} \\
    \midrule
    Raw ensemble & -- & 1.0619 & 1.8041 \\
    BIV-EMOS & $1.3607 \pm 0.0019$ & $0.4493 \pm 0.0003$ & $0.7581 \pm 0.0009$ \\
    BIV-DRN & $1.1734 \pm 0.0227$ & $0.4283 \pm 0.0070$ & $0.7109 \pm 0.0201$ \\
    DCDN & $0.4742 \pm 0.0263$ & $0.3661 \pm 0.0046$ & $0.5493 \pm 0.0094$ \\
    Improvement of BIV-DRN (\%) & $20.63 \pm 2.70$ & $4.68 \pm 1.52$ & $6.23 \pm 2.58$ \\
    Improvement of DCDN (\%) & $142.74 \pm 6.55$ & $18.51 \pm 1.05$ & $27.55 \pm 1.28$ \\
    \bottomrule
  \end{tabular}
\end{table}

\subsection{Sites and lead times}\label{sec:sites}

\begin{figure}[t]
  \centering
  \includegraphics{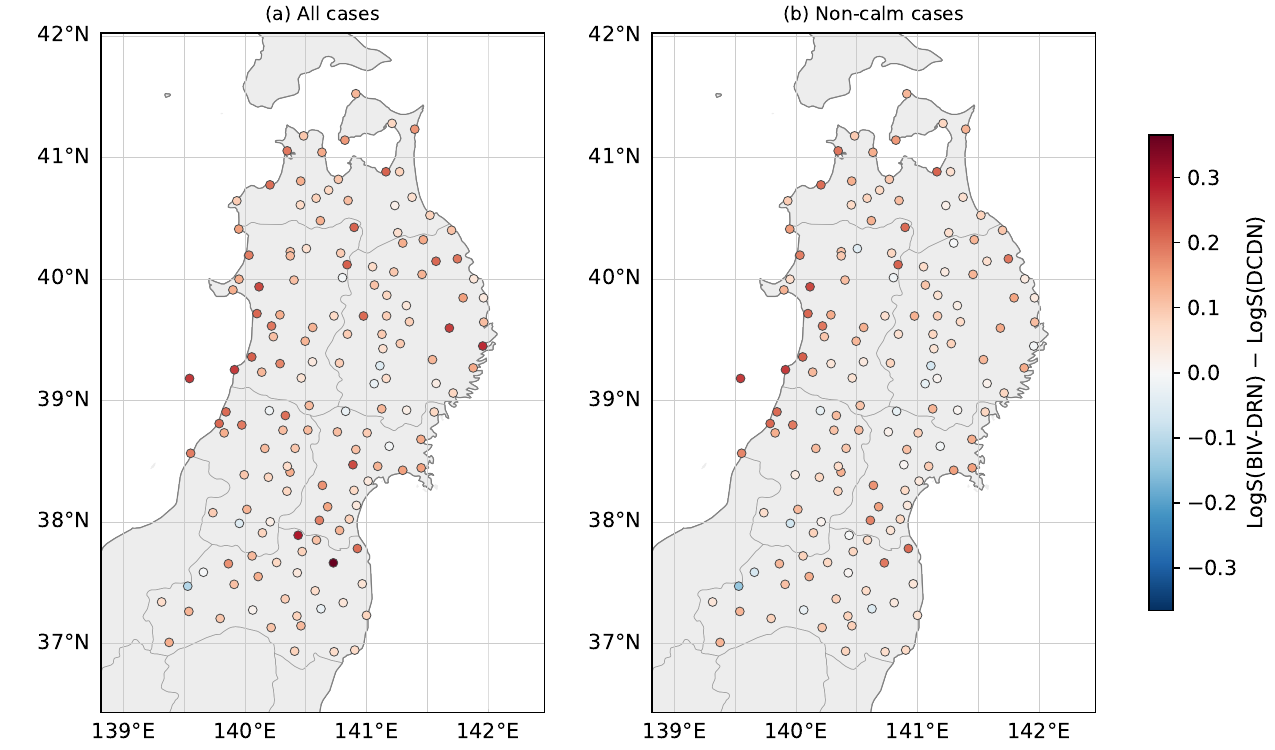}
  \caption{Improvement of DCDN over BIV-DRN at each site, given as the LogS of BIV-DRN minus that of DCDN averaged over (a) all test cases and (b) non-calm test cases. Positive values favor DCDN}
  \label{fig:sitemap}
\end{figure}

\begin{figure}[t]
  \centering
  \includegraphics{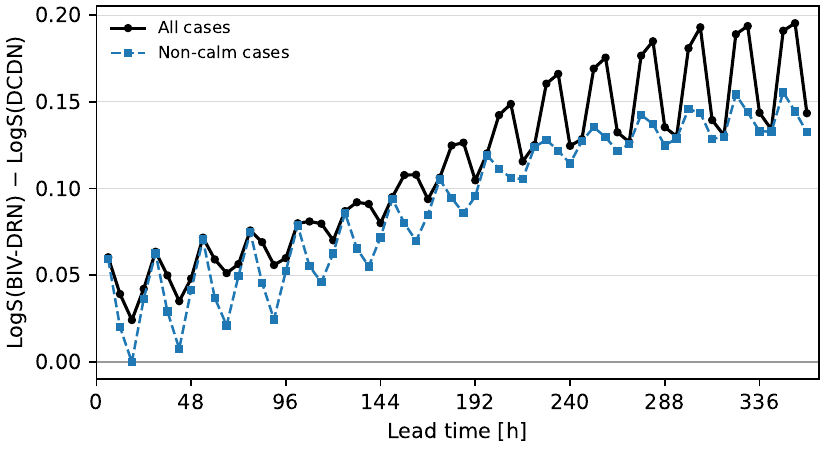}
  \caption{As \Fig{fig:sitemap}, but for each lead time, for all test cases (solid) and non-calm test cases (dashed)}
  \label{fig:leadtime}
\end{figure}

The improvement in LogS of DCDN over BIV-DRN is shown for each site in \Fig{fig:sitemap} and for each lead time in \Fig{fig:leadtime}.
This improvement is the LogS of BIV-DRN minus that of DCDN, averaged over the cases at each site or lead time.
Each improvement is shown for all cases and for the non-calm cases.

The improvement was positive at 144 of the 154 sites in all cases and at 139 sites in the non-calm cases (\Fig{fig:sitemap}).
In all cases, the median improvement across the sites was 0.108, and the improvement ranged from $-0.108$ to 0.365.
In the non-calm cases, the median was 0.086, and the improvement ranged from $-0.141$ to 0.256.
In the non-calm cases, many of the sites with the largest improvements were near the coast of the Sea of Japan.
In both sets of cases, the sites with negative improvements were scattered.

The improvement increased with lead time in all cases and in the non-calm cases (\Fig{fig:leadtime}).
In all cases, it was positive at all 60 lead times and increased from 0.060 at 6~h to 0.143 at 360~h.
In the non-calm cases, it was positive at all lead times except 18~h ($-0.0003$) and increased from 0.059 at 6~h to 0.132 at 360~h.
The daily oscillation of the improvement was larger in all cases than in the non-calm cases.

\subsection{Observed wind speeds}\label{sec:speed}

\begin{figure}[t]
  \centering
  \includegraphics{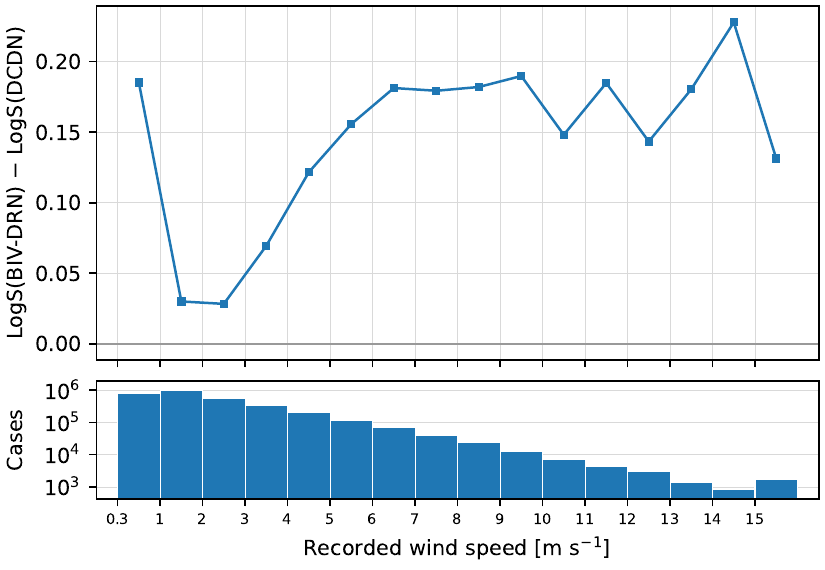}
  \caption{As \Fig{fig:sitemap}, but for each interval of the recorded wind speed of the non-calm test cases. The last interval contains all cases of $15\ms$ or more. Lower panel: number of cases}
  \label{fig:speed}
\end{figure}

The improvement in LogS of DCDN over BIV-DRN in the non-calm cases is shown for each interval of the recorded wind speed in \Fig{fig:speed}.
The non-calm cases are grouped into intervals of $1\ms$, with the cases of $15\ms$ or more merged.
As in \Fig{fig:sitemap}, the improvement is the LogS of BIV-DRN minus that of DCDN averaged over the cases in each interval.

The improvement was large near zero wind speed and above about $5\ms$, and small between 1 and $3\ms$.
The mean improvement was 0.699 in the calm cases (\Tab{tab:calm}) and 0.185 in the non-calm cases below $1\ms$ (\Fig{fig:speed}).
Between 1 and $3\ms$, it was about 0.03, the smallest of all intervals.
Above $3\ms$, it increased with wind speed and was between 0.13 and 0.23 in every interval from $5\ms$ upward.
As with \Tab{tab:calm}, this breakdown by the observations is descriptive.

\subsection{Transformed observations}\label{sec:pit-results}

\begin{figure}[t]
  \centering
  \includegraphics{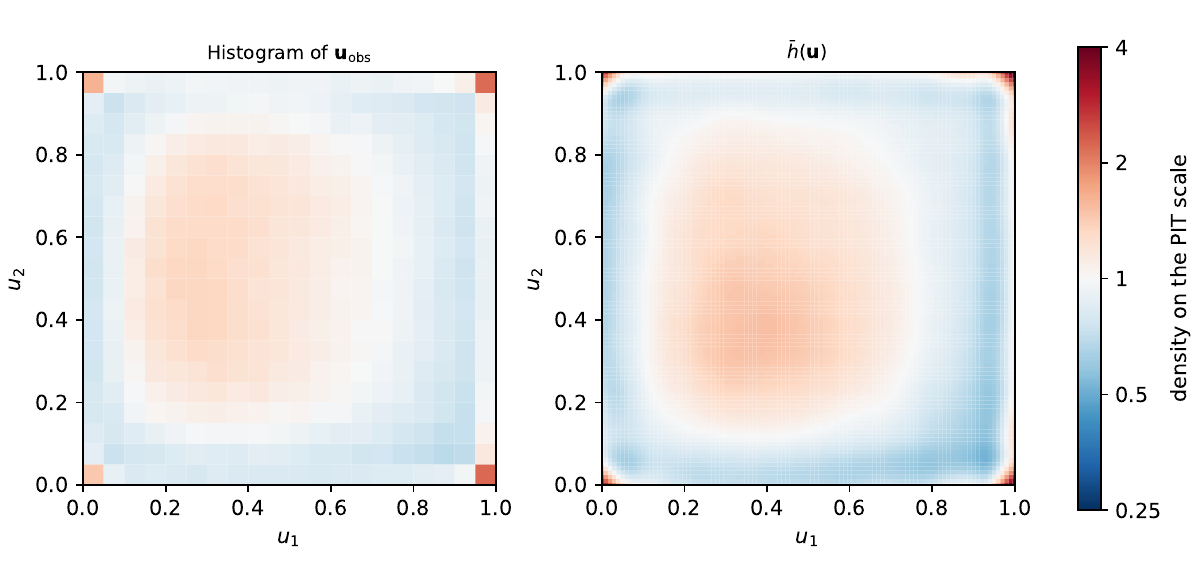}
  \caption{Test period on the PIT scale. Left: histogram of the transformed observations, normalized as a density. Right: average correction density $\bar h$ of DCDN. Values above 4 are shown in the darkest red}
  \label{fig:pit}
\end{figure}

\Fig{fig:pit} compares the two-dimensional PIT histogram of the transformed observations with the uniform density and the average correction density.
The transformed observation $\mathbf{u}_{\mathrm{obs}}$ is the dequantized observation mapped by $T$ of BIV-DRN, and the histogram is normalized as a density.
If the observations follow BIV-DRN, $\mathbf{u}_{\mathrm{obs}}$ is uniformly distributed on the PIT scale.
If they follow DCDN, the density of $\mathbf{u}_{\mathrm{obs}}$ in each case is the correction density $h$ of that case (\Sec{sec:dcdn}).
All cases together then follow the mixture of these densities, which equals the average $\bar h$ of $h$ over all test cases.

The histogram of the transformed observations departed from the uniform density, and the average correction density showed the same main features (\Fig{fig:pit}).
Both exceeded one in the central part of the PIT scale, with the peak shifted toward small $u_1$.
Both were below one in a band along the edges and exceeded one at the four corners.

\section{Discussion and conclusions}\label{sec:discussion}

DCDN improved on BIV-DRN in all three scores over the test period at 154 sites in the Tohoku region.
Relative to BIV-EMOS, DCDN improved the LogS, the ES, and the VS by 18.9\%, 3.3\%, and 4.0\%, respectively.
BIV-DRN improved them by 6.0\%, 2.5\%, and 3.3\%.
All nine DM tests were significant.
The improvement in LogS was positive at most sites and lead times, also in the non-calm cases.

The results are consistent with the interpretation that the bivariate normal distribution of BIV-DRN misses part of the conditional distribution of the observed wind vector.
The transformed observations in \Fig{fig:pit} departed from the uniform density expected under BIV-DRN, and the average correction density had the same features.
With 8 instead of 13 B-splines in each coordinate, DCDN was slightly worse in all three scores, whereas with 23 it improved further only in the LogS.
In another variant, the correction was locally restricted to directions whose score functions at zero correction are Fisher orthogonal to those of the reference mean and covariance.
This variant retained 96\%, 94\%, and 81\% of the improvement of DCDN over BIV-DRN in the LogS, ES, and VS, respectively.

The main features of the transformed observations are also consistent with specific non-normal features of the conditional distribution.
The central peak and the high corners mean that, compared with BIV-DRN, more observations fell near its mean and where both whitened components were extreme.

The improvement in LogS by observed wind speed is also consistent with a conditional distribution concentrated near zero with a relatively heavy tail toward higher wind speeds.
The improvement was large near zero wind speed and above about $5\ms$, and small between 1 and $3\ms$ (\Sec{sec:speed}).
For each case, a positive improvement means that DCDN gave the observation a higher density than BIV-DRN, because DCDN multiplies the BIV-DRN density by the correction density.
Near zero and at higher wind speeds, DCDN thus gave the observations a higher log density than BIV-DRN on average.
The improvement at higher observed wind speeds may be relevant to applications such as wind energy \citep{schuhen2012ensemble}.

The results are limited by the compared methods, the data, and an unverified dequantization assumption.
Only BIV-DRN was tested as the reference density, and the comparison excludes the flexible methods of \citet{buchner2026bivariate}.
The evaluation covers a single region and a single year, 2025.
The ECMWF ensemble prediction system was updated several times from the training period to the test period.
The dequantization assumes that the wind vector is uniform within the set sharing each recorded value, which the discrete records cannot confirm (\Sec{sec:dequantization}).
The improvement in the VS is the least certain: it came from the calm cases, and its DM test was not adjusted for the nine comparisons.

DCDN can correct a bivariate normal reference density while containing it as a special case.
Because the reference density is kept fixed, the correction can be trained on top of an existing BIV-DRN without retraining it.
The correction improved the LogS not only near zero wind speed but also at higher observed wind speeds.
This is consistent with non-normal features that the bivariate normal distribution misses.
Correcting the bivariate normal distribution, rather than replacing it, can thus add flexibility while improving the scores.

\paperappendix

\section{Implementation details}\label{app:details}

\subsection{Inputs}\label{app:inputs}

The ensemble statistics in \Tab{tab:inputs} are computed from the $K = 50$ perturbed members $e_{k,c}$ of each wind component $c \in \{u, v\}$, not from the control forecast.
For each component, they are the mean $\bar e_c$, the standard deviation $s_c$, the skewness $\gamma_c$, and the excess kurtosis $\kappa_c$:
\begin{gather*}
  \bar e_c = \frac{1}{K} \sum_{k=1}^{K} e_{k,c}, \qquad
  s_c = \biggl( \frac{1}{K-1} \sum_{k=1}^{K} (e_{k,c} - \bar e_c)^2 \biggr)^{1/2}, \\
  \gamma_c = \frac{m_{3,c}}{m_{2,c}^{3/2}}, \qquad
  \kappa_c = \frac{m_{4,c}}{m_{2,c}^{2}} - 3, \qquad
  m_{n,c} = \frac{1}{K} \sum_{k=1}^{K} (e_{k,c} - \bar e_c)^n .
\end{gather*}
The input for the standard deviation is $\ln s_c$.
The correlation between the two components is the sample correlation coefficient
\begin{equation*}
  r = \frac{\sum_{k=1}^{K} (e_{k,u} - \bar e_u)(e_{k,v} - \bar e_v)}
  {\bigl( \sum_{k=1}^{K} (e_{k,u} - \bar e_u)^2 \sum_{k=1}^{K} (e_{k,v} - \bar e_v)^2 \bigr)^{1/2}} .
\end{equation*}

The remaining inputs are the control forecast, the lead time, and the time of day and the day of the year.
The control forecast of each component is used as it is, and the lead time is given in hours.
The time of day and the day of the year are taken at the valid time.
Let $t$ be the time of day in hours UTC, and let $d$ be the day of the year, with $d = 1$ on 1 January.
Their inputs are
\begin{equation*}
  \sin \frac{2\pi t}{24}, \qquad \cos \frac{2\pi t}{24}, \qquad
  \sin \frac{2\pi d}{365.25}, \qquad \cos \frac{2\pi d}{365.25} .
\end{equation*}

\subsection{Sampling from DCDN}\label{app:sampling}

Samples from DCDN are drawn by choosing a mixture component, drawing each coordinate from its normalized B-spline, and inverting $T$.
First, a component $(i, j)$ is chosen with probability $p_{ij}$.
Second, $u_1$ and $u_2$ are drawn independently as averages of the five consecutive knots of $B_i$ and $B_j$, respectively, with weights drawn uniformly from the simplex.
The average of the knots of $B_i$ with these weights has the density $\phi_i$ \citep{karlin1986multivariate}.
Finally, $T$ is inverted to give the sample
\begin{equation*}
  \mathbf{z} = \Phi^{-1}(\mathbf{u}), \qquad
  \mathbf{y} = \boldsymbol{\mu} + \boldsymbol{\Sigma}^{1/2}\mathbf{z} .
\end{equation*}
The sample $\mathbf{y}$ therefore follows $f$.

\subsection{Training and random numbers}\label{app:training}

The random seed sets the initial weights, the order of the training batches, and the perturbations of the training observations (\Sec{sec:dequantization}).
It also sets the random numbers for the 1000 samples of the ES and the VS.
The ten training runs of each method use the seeds 0 to 9.
The perturbations of the validation and test observations use a fixed random seed of 0 for all training runs.
Training and evaluation use deterministic computations, so that differences between seeds arise only from these random numbers.
The standard deviations of the ES and the VS in Tables~\ref{tab:overall} and~\ref{tab:calm} therefore also include sampling variation.

Models of different methods are paired by seed.
DCDN with each seed uses BIV-DRN with the same seed as its reference density, and each improvement pairs the models with the same seed.
The preliminary search for the hyperparameters used seed 0, and the fixed hyperparameters are used for all seeds.
The DM tests and the results shown in figures use the models with seed 0.

\bibliographystyle{apalike-arxiv}
\bibliography{references}

\section*{Acknowledgments}
The first author was supported by the Morinokuni Scholarship Foundation and the Shigemitsu Sunaga Fund at Tohoku University.

This work is based on TIGGE data.
TIGGE (The International Grand Global Ensemble) is an initiative of the World Weather Research Programme (WWRP).

The first author used Claude Code (Anthropic) and Codex (OpenAI) to write the code and draft the manuscript.
The first author reviewed the generated code and text and takes full responsibility for the content of this paper.
\section*{Data Availability}
\defcitealias{ecmwf2026tigge}{ECMWF, 2026}
The ensemble forecasts are based on data and products of the European Centre for Medium-Range Weather Forecasts (ECMWF), obtained from the TIGGE dataset (\citetalias{ecmwf2026tigge}; \citealp{bougeault2010thorpex,swinbank2016tigge}).
\copyright{} 2026 ECMWF.
Source: ECMWF Data Store at ecds.ecmwf.int.
This data is published under a Creative Commons Attribution 4.0 International (CC BY 4.0) licence (\url{https://creativecommons.org/licenses/by/4.0/legalcode}).
ECMWF does not accept any liability whatsoever for any error or omission in the data, their availability, or for any loss or damage arising from their use.
We extracted the forecasts at the grid points nearest to the sites.

The wind observations are based on ``Past weather data download'' (Japan Meteorological Agency) (\url{https://www.data.jma.go.jp/risk/obsdl/index.php}).

\end{document}